\documentclass[3p,times,procedia]{elsarticle}
\usepackage{nupha_ecrc}
\usepackage{wrapfig}
\volume{00}
\firstpage{1}
\journalname{Nuclear Physics A}
\runauth{G. Vujanovic et al.}
\jid{nupha}
\jnltitlelogo{Nuclear Physics A}
\usepackage{amssymb}
\usepackage[figuresright]{rotating}
\begin{document}
\begin{frontmatter}
\dochead{XXVIth International Conference on Ultrarelativistic Nucleus-Nucleus Collisions\\ (Quark Matter 2017)}

\title{Bulk viscous effects on flow and dilepton radiation\newline in a hybrid approach}
\author[label1]{Gojko Vujanovic}
\address[label1]{Department of Physics, The Ohio State University, 191 West Woodruff Avenue, Columbus, Ohio 43210, USA}
\author[label2]{Jean-Fran\c cois Paquet}
\address[label2]{Department of Physics \& Astronomy, Stony Brook University, Stony Brook, New York 11794, USA }
\author[label3,label6]{Sangwook Ryu}
\address[label3]{Frankfurt Institute for Advanced Studies, Ruth-Moufang-Str. 1, 60438, Frankfurt am Main, Germany}
\author[label4]{Chun Shen}
\address[label4]{Department of Physics, Brookhaven National Laboratory, Upton, New York 11973-5000}
\author[label5]{Gabriel S. Denicol}
\address[label5]{Instituto de F\'{i}sica, Universidade Federal Fluminense, UFF, Niter\'{o}i, 24210-346, RJ, Brazil}
\author[label6]{Sangyong Jeon}
\author[label6]{Charles Gale}
\address[label6]{Department of Physics, McGill University, 3600 University Street, Montr\'eal, QC, H3A 2T8, Canada}
\author[label1]{Ulrich Heinz}

\begin{abstract}
Starting from IP-Glasma initial conditions, we investigate the effects of bulk pressure on low mass dilepton production at the Relativistic Heavy Ion Collider (RHIC) and the Large Hadron Collider (LHC) energies. Though thermal dilepton $v_2$ is affected by the presence of both bulk and shear viscosity, whether or not these effects can be measured depends on the dilepton ``cocktail'' contribution to the the low mass dilepton $v_2$. Combining the thermal and ``cocktail'' dileptons, the effects of bulk viscosity on total dilepton $v_2$ is investigated. 
\end{abstract}


\end{frontmatter}

\section{Introduction}\label{sec:intro}
The study of the properties of the strongly interacting QCD medium has been at the center of modern relativistic heavy-ion experiments. At the heart of these investigations is the characterization of the medium's in-equilibrium properties, given by its equation of state --- which has recently been computed using lattice QCD ($\ell$QCD) \cite{Borsanyi:2013bia,Bazavov:2014pvz} --- as well as its near-equilibrium transport properties captured by transport coefficients such as shear ($\eta$) and bulk ($\zeta$) viscosities. A significant effort into determining these properties has recently been undertaken using Bayesian statistics, for a comparison between soft hadronic experimental observables and theoretical calculations, which has revealed some promising constraints. For example, Ref. \cite{Pratt:2015zsa} shows that a parametrization of the speed of sound close to that of $\ell$QCD calculations best describes experimental data of soft hadronic observables, thus these observables have potential to constrain the equation of state qualitatively. In the same vein, another recent Bayesian analysis has produced limits on the temperature-dependences of $\eta$ \cite{Bernhard:2016tnd} and $\zeta$ \cite{Bernhard:QM2017}. However, soft hadronic observables are not evenly sensitive to all transport coefficients of QCD. Indeed, studies \cite{Song:2008si, Vujanovic:2016anq} show that hadrons are less sensitive to the relaxation time ($\tau_\pi$) of shear pressure than to $\eta$, for example. Electromagnetic (EM) probes, on the other hand, do show an increased sensitivity to $\tau_\pi$ \cite{Vujanovic:2016anq}, while also being quite sensitive to other transport coefficients \cite{Vujanovic:2015nwv,Vujanovic:2017psb,Paquet:2015lta,Vujanovic:2017fkh}; thus hadronic and EM observables complement each other. This contribution focuses on a particular subcategory of EM radiation, namely lepton pair (or dilepton) emission. Our goal is further enhance our understanding of the sensitivity of dileptons to transport coefficients, in particular to bulk viscosity.

\section{Dilepton production}\label{sec:dilepton_rates}
We consider two categories of dilepton radiation. The first is the thermal emission originating from the hydrodynamical evolution of the medium itself. The second source are the ``cocktail'' dileptons, composed of late direct decays of vector mesons and Dalitz decays of mesons into lepton pairs \cite{Landsberg:1986fd}. 

At leading order in the electromagnetic coupling $\alpha$, and to all orders in the QCD coupling $\alpha_s$ the thermal dilepton rates can be written as:
\begin{eqnarray}
\frac{d^4 R}{d^4 p} = -\frac{\alpha}{12\pi^4} \frac{1}{M^2} \frac{{\rm Im} \Pi_{{\rm R},\,\gamma^{\ast}}}{\exp\left[p\cdot u/T\right]-1},
\label{eq:rate}
\end{eqnarray}
where $M^{2}=p^\mu p_\mu$ is the virtual photon invariant mass, $p^\mu$ is the virtual photon four momentum, $u^\mu$ is the four velocity of the emitting medium, while $T$ is its temperature --- $u^\mu$ and $T$ are obtained by solving the hydrodynamical equations of motion, see below --- and $\Pi_{{\rm R},\,\gamma^{\ast}} = g_{\mu\nu}\Pi^{\mu\nu}_{{\rm R},\,\gamma^{\ast}}$ is the trace of the retarded virtual photon self-energy. 

The thermal dilepton rate is computed by combining rates using hadronic and partonic degrees of freedom. The partonic dilepton emission rate stems from a calculation of quark-antiquark annihilation in the Born approximation. This dilepton rate is directly proportional to the quark/antiquark distribution functions \cite{Vujanovic:2013jpa}, that, in a dissipative medium, acquire viscous corrections accounting for the deformation viscosity induces on the thermal equilibrium momentum distribution. The correction related to the shear tensor is computed using the 14-moment approximation first proposed by Israel and Stewart \cite{Israel1976310,Israel:1979wp}, whereas the correction arising from to bulk viscous pressure is calculated following the method described in Refs. \cite{Paquet:2015lta,Tinti:2016bav}. In the hadronic medium (HM), the major source of dileptons originates from in-medium decay of vector mesons, which is well described by the Vector Meson Dominance (VMD) model. The hadronic  dilepton rate in VMD is proportional to the imaginary part of the in-medium vector meson $(V)$ propagator ${\rm Im}\left[D_{V}^{\mathrm{R}}\right]$, which in turn requires the knowledge of the self-energy $\Pi _{V}$ of vector mesons in the medium. Of course, $\Pi _{V}$ includes dissipative contributions. The inviscid rate and its shear viscous corrections are discussed in detail in Ref. \cite{Vujanovic:2013jpa}. Bulk viscous corrections to the thermal distribution function are computed using results derived in Refs. \cite{Paquet:2015lta,Jaiswal:2014isa}. The interpolation between the hadronic and partonic degrees of freedom is done using a switching function that is linear in the QGP fraction inside a space-time cell. Consistent with the employed equation of state \cite{Huovinen:2009yb}, this linear interpolation is performed within the interval $184<T<220$ MeV. Thermal dilepton emission was only evaluated above the switching temperature, see discussion below. 

Lastly, the dilepton cocktail is computed using two different approaches. The first solely uses a hydrodynamical calculation to obtain the final hadronic spectrum; this is done via the Cooper-Frye formalism including hadronic resonance decays. The second approach dynamically evolves hadrons below the switching surface with UrQMD, without letting those hadrons decay into dileptons during UrQMD evolution, thus capturing additional hadronic dynamics modifying the hadron momentum distributions to be decayed into dileptons. In both cases, the final hadron spectrum is decayed into dileptons through VMD \cite{Landsberg:1986fd}.

\section{Initial conditions and viscous hydrodynamics}\label{sec:visc_hydro}
The hydrodynamical evolution equations are initialized with IP-Glasma initial conditions \cite{Paquet:2015lta,Ryu:2015vwa}. The dynamical evolution of the fluid consists of energy-momentum conservation $\partial_\mu T^{\mu\nu}=0$, and relaxation equations for the dissipative bulk pressure ($\Pi$) and shear tensor ($\pi^{\mu\nu}$). The energy-momentum tensor is decomposed as $T^{\mu\nu}=\varepsilon u^\mu u^\nu - (P+\Pi)\Delta^{\mu\nu}+\pi^{\mu\nu}$, where $\varepsilon$ is the energy density, $u^\mu$ is the flow four velocity, $P$ is the thermodynamic pressure related to $\varepsilon$ by the equation of state $P(\varepsilon)$ \cite{Huovinen:2009yb}, $\Delta^{\mu\nu}=g^{\mu\nu}-u^\mu u^\nu$, and $g^{\mu\nu}={\rm diag}(+,-,-,-)$ is the metric tensor. The relaxation equations for $\Pi$ and $\pi^{\mu\nu}$ are:
\begin{eqnarray}
\tau_\Pi \dot{\Pi}+\Pi &=& -\zeta\theta \nonumber - \delta_{\Pi\Pi}\Pi\theta + \lambda_{\Pi\pi}\pi^{\alpha\beta}\sigma_{\alpha\beta},\\
\tau_\pi \dot{\pi}^{\langle\mu\nu\rangle}+\pi^{\mu\nu} &=& 2\eta\sigma^{\mu\nu}-\delta_{\pi\pi}\pi^{\mu\nu}\theta + \lambda_{\pi\Pi}\Pi\sigma^{\mu\nu} - \tau_{\pi\pi} \pi^{\langle\mu}_\alpha\sigma^{\mu\rangle\alpha} + \phi_7 \pi^{\langle\mu}_\alpha\pi^{\mu\rangle\alpha},
\label{eq:bulk_shear}
\end{eqnarray}  
where $\dot{\Pi}=u^\alpha\partial_\alpha\Pi$, $\dot{\pi}^{\langle\mu\nu\rangle}=\Delta^{\mu\nu}_{\alpha\beta}u^\lambda\partial_\lambda\pi^{\alpha\beta}$, $\Delta_{\alpha\beta}^{\mu\nu}=\left(\Delta_{\alpha}^{\mu}\Delta_{\beta}^{\nu}+\Delta_{\beta}^{\mu}\Delta_{\alpha}^{\nu}\right)/2-(\Delta_{\alpha\beta}\Delta^{\mu\nu})/3$, $\theta=\partial_\alpha u^\alpha$, and $\sigma^{\mu\nu}=\partial^{\langle\mu} u^{\nu\rangle}$. The specific shear viscosity $\eta/s$ --- where $s$ is the entropy density --- is assumed to be almost temperature independent, while the specific bulk viscosity ($\zeta/s$) is temperature-dependent as shown in Refs. \cite{Paquet:2015lta,Ryu:2015vwa}. Those references also contains further details about the prescription used to fix the second-order transport coefficients in Eqs. (\ref{eq:bulk_shear}). A weak temperature dependence of $\eta/s$ is introduced by permitting for $\eta/s$ to change slightly between collision energies, going from top RHIC energy of $\sqrt{s_{NN}}=200$ GeV to LHC energy of $\sqrt{s_{NN}}=2.76$ TeV. Hydrodynamical simulations are evolved until the switching temperature ($T_{sw}$) is reached, where hydrodynamical degrees of freedom are converted to hadrons. Further hadronic dynamics of the medium are performed using UrQMD simulations. The switching temperature at which the best description of the hadronic observables at top RHIC collision energy is reached was found to be $T^{RHIC}_{sw}=165$ MeV, whereas at LHC energy a temperature $T^{LHC}_{sw}=145$ MeV \cite{Paquet:2015lta,Ryu:2015vwa} is needed. 

\section{Results}\label{sec:results}
\vspace{-0.5cm}
\begin{figure}[!h]
\includegraphics[width=0.445\textwidth]{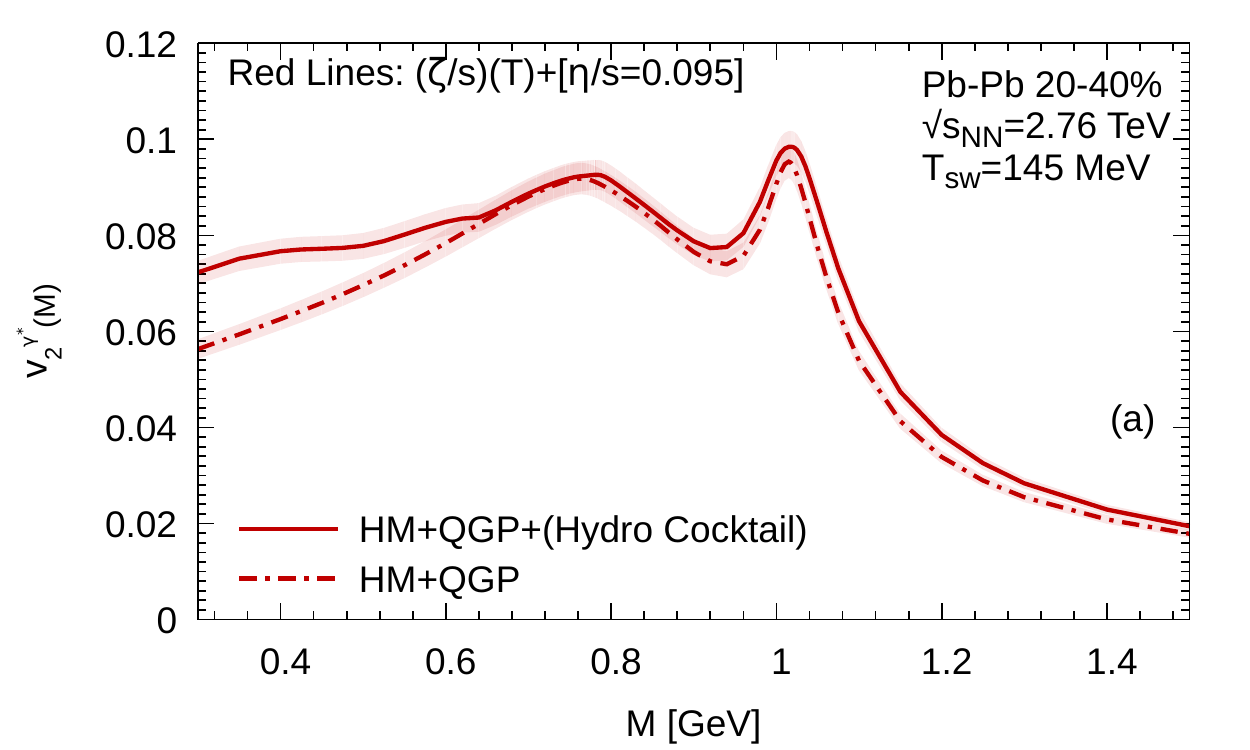} 
\includegraphics[width=0.445\textwidth]{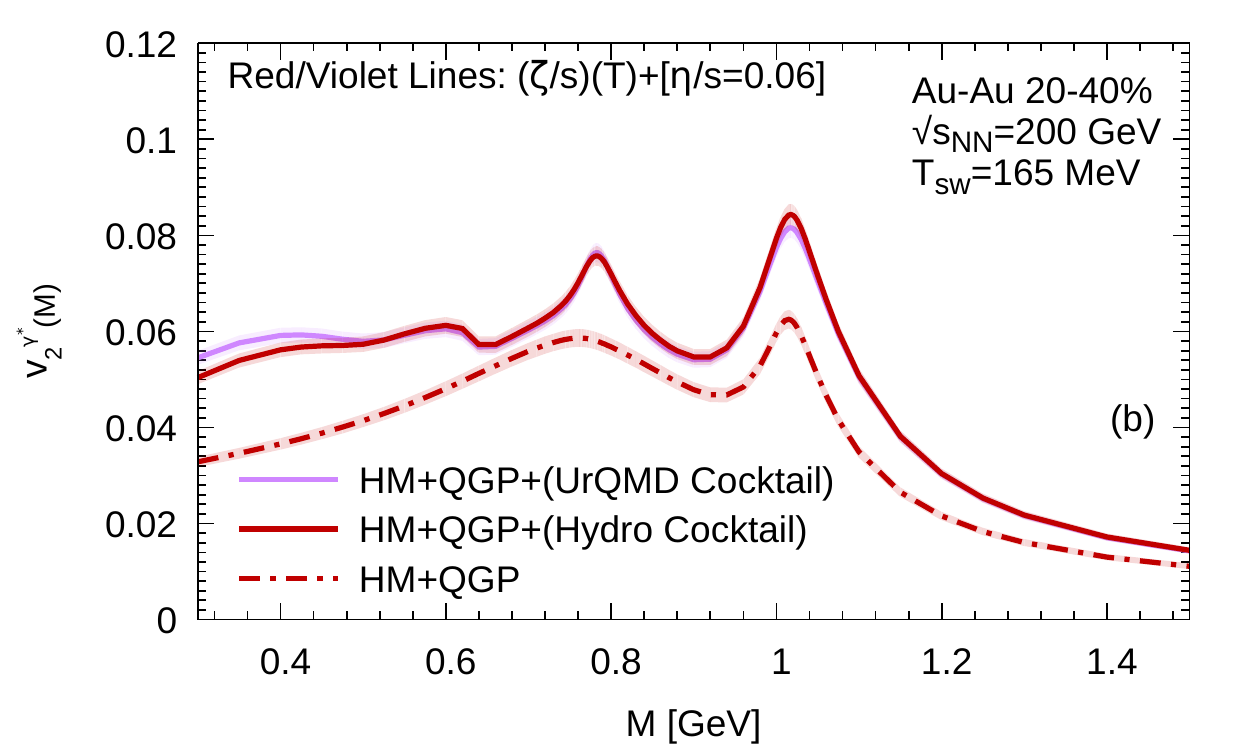}
\caption{(Color online) (a) Dilepton $v_2(M)$ at LHC energy with and without cocktail dileptons. Hydro Cocktail is obtained from the Cooper-Frye formalism with resonance decays. (b) Dilepton $v_2(M)$ at top RHIC energy with/without cocktail dileptons.}
\label{fig:hydro_vs_urqmd_cocktail}
\end{figure} 
The importance of the dilepton cocktail is explored in Fig. \ref{fig:hydro_vs_urqmd_cocktail}. Owing to the larger space-time volume described by the hydrodynamical evolution at LHC energy compared to top RHIC energy, more thermal dileptons will be radiated at LHC than at RHIC. Since this relative increase in space-time volume at LHC persists also to lower temperatures (as $T^{LHC}_{sw}<T^{RHIC}_{sw}$), it is not surprising that most of the total $v_2(M)$ of dileptons at the LHC is coming from the hydrodynamical evolution itself. Cocktail dileptons are important only for $M<0.65$ GeV. At top RHIC energy, the overall space-time volume that is evolved hydrodynamically is smaller, thus cocktail dileptons leave a larger imprint on the overall dilepton $v_2(M)$, as depicted in Fig. \ref{fig:hydro_vs_urqmd_cocktail}b. However, at top RHIC energy, differences between the two models employed to obtain the dilepton cocktail appear to be less relevant. Indeed, both UrQMD and a purely hydrodynamical calculation give rise to a similar final $v_2(M)$. The dynamical change caused by UrQMD of the momentum spectra of hadrons decaying into dileptons does not appear to have a large effect on the dilepton elliptic flow. To determine the role of a dynamical generation of cocktail dileptons, dilepton production during a hadronic transport evolution will be explored in the future.      

\begin{figure}
\includegraphics[width=0.445\textwidth]{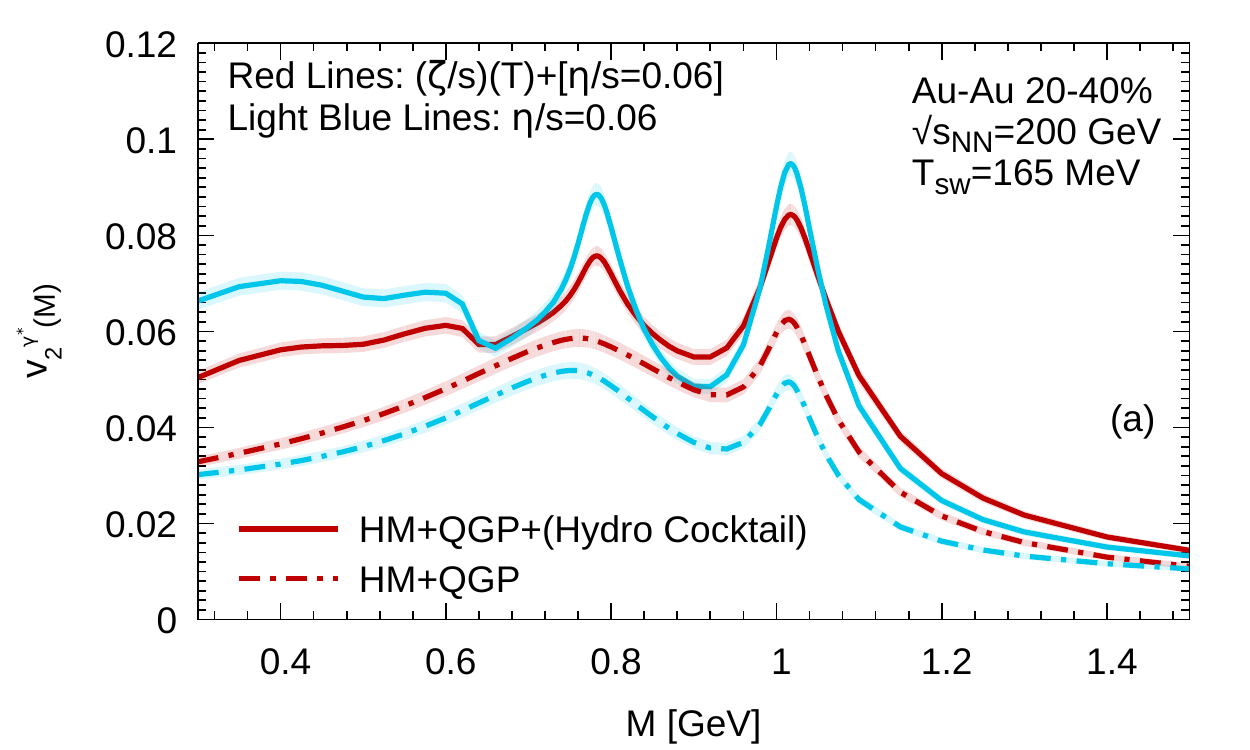}
\includegraphics[width=0.445\textwidth]{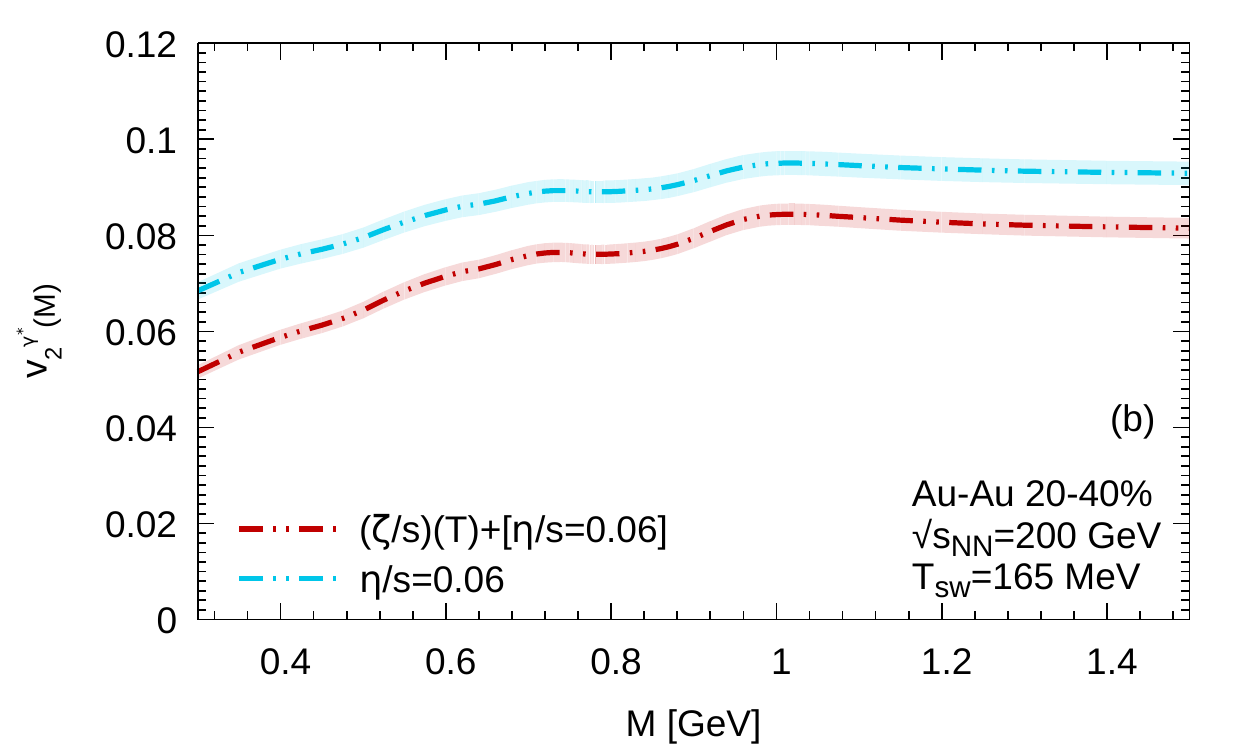}
\caption{(Color online) Bulk viscous effects on the full (thermal + cocktail) and thermal dilepton $v_2(M)$ (a) and on cocktail dilepton $v_2(M)$ (b).}
\label{fig:bulk_vs_shear_RHIC}
\end{figure}


At top RHIC energy, the effects of bulk viscosity relative to shear viscosity on the total $v_2(M)$ are presented in Fig. \ref{fig:bulk_vs_shear_RHIC}. In the invariant mass windows $0.85<M<0.95$ GeV and $M>1.1$ GeV, the ordering of the blue and red solid curves changes relative to other invariant masses. This effect is caused by a competition between thermal and cocktail dilepton $v_2$. Indeed, the thermal dilepton $v_2$ in Fig. \ref{fig:bulk_vs_shear_RHIC}a shows that a medium with bulk viscosity increases thermal dilepton $v_2$ while suppressing the dilepton cocktail $v_2$ see Fig. \ref{fig:bulk_vs_shear_RHIC}b. The reasons why bulk viscosity causes an increase in flow anisotropy of thermal dileptons will be investigated further in an upcoming publication.  

\section{Conclusion}\label{sec:conc}
In this contribution, we have explored the effect of bulk viscosity and the dilepton cocktail on the overall dilepton $v_2$. Indeed, thermal dileptons are highly sensitive to the details of the hydrodynamical evolution both at RHIC and at LHC energies. The footprint of the dilepton cocktail is more significant at top RHIC energy than it is at the LHC energies. A more in-depth study about the effects of bulk viscosity, and of the importance of the dilepton cocktail on the total $v_2$, will be performed in the near future. 

\section*{Acknowledgements}
This work was supported in part by the Natural Sciences and Engineering Research Council of Canada, by the Director, Office of Energy Research, Office of High Energy and Nuclear Physics, Division of Nuclear Physics, of the U.S. Department of Energy under awards No. \rm{DE-SC0004286}, \rm{DE-FG02-88ER40388}, and by the National Science Foundation (in the framework of the JETSCAPE Collaboration) through award No. 1550233. G. V., G. S. D., and  C. G. gratefully acknowledge support by the Fonds de Recherche du Qu\'ebec --- Nature et les Technologies (FRQ-NT), through the Banting Fellowship from the Government of Canada, and from the Canada Council for the Arts through its Killam Research Fellowship program, respectively. Computations were performed on the Guillimin supercomputer at McGill University under the auspices of Calcul Qu\'ebec and Compute Canada. The operation of Guillimin is funded by the Canada Foundation for Innovation (CFI), the National Science and Engineering Research Council (NSERC), NanoQu\'ebec, and the Fonds de Recherche du Qu\'ebec --- Nature et les Technologies (FRQ-NT).

\nocite{*}
\bibliographystyle{elsarticle-num}
\bibliography{references_short}
\end{document}